\documentclass[a4paper,10pt,twoside]{cpc-hepnp}

\usepackage{multicol}
\usepackage{graphicx}
\usepackage{booktabs}
\usepackage{amssymb,bm,mathrsfs,bbm,amscd}
\usepackage[tbtags]{amsmath}
\usepackage{lastpage}

\begin{document}

\fancyhead[c]{\small Chinese Physics C~~~Vol. xx, No. x (201x) xxxxxx}
\fancyfoot[C]{\small 010201-\thepage}

\footnotetext[0]{Received 16 February 2015}

\title{Drive laser system for the DC-SRF photoinjector at Peking University\thanks
{Supported by National Basic Research Project (973) 
(No. 2011CB808302  and No. 2011CB808304)}}

\author{%
      Zhiwen Wang
\quad Senlin Huang$^{1)}$\email{huangsl@pku.edu.cn}%
\quad Lin Lin
\quad Gang Zhao
\quad Shengwen Quan
\quad Kexin Liu
\quad Jiaer Chen
}
\maketitle

\address{%
State Key Laboratory of Nuclear Physics and Technology, Institute of Heavy Ion Physics, Peking University, Beijing 100871, China\\
}

\begin{abstract}
Photoinjectors are widely used for linear accelerators as electron sources to generate high-brightness electron beam. Drive laser, which determines the timing structure and quality of the electron beam, is a crucial device of photoinjector. A new drive laser system has been designed and constructed for the upgraded 3.5-cell DC-SRF photoinjector at Peking University. The drive laser system consists of a 1064 nm laser oscillator, a four-stage amplifier, the second and fourth harmonic generators, the optical system to transfer the UV pulses to the photocathode, and the synchronization system. The drive laser system has been successfully applied in the stable operation of DC-SRF photoinjector and its performance meets the requirements. 266 nm laser with an average power close to 1W can be delivered to illuminate the Cs${}_2$Te photocathode and the instability is less than $5\%$ for long time operation. The design consideration for improving the UV laser quality, a detailed description of laser system, and its performance are presented in this paper. \end{abstract}



\footnotetext[0]{\hspace*{-3mm}\raisebox{0.3ex}{$\scriptstyle\copyright$}2013
Chinese Physical Society and the Institute of High Energy Physics
of the Chinese Academy of Sciences and the Institute
of Modern Physics of the Chinese Academy of Sciences and IOP Publishing Ltd}%

\begin{multicols}{2}

\section{Introduction}

Photoinjectors are widely used for linear accelerators as electron sources to generate high-brightness electron beam after the development of more than two decades. Except determining the basic properties of the electron bunches, such as their time sequence, duration, and the amount of bunch charge, driver laser also affects the stability of the produced bunch and its synchronization accuracy to the accelerator RF field, as well as the emmitance of the electron beam. As a crucial device of photoinjector, drive laser systems are generally developed by many laboratories themselves because there are few commercially available products which can meet the requirements of different applications. For example, for the ERL-FEL at JLab, a drive laser with the pulse repetition rate of $75$ MHz, wavelength of $532$ nm, and average power of $25$ W has been developed~\cite{ZhangS2009}. For the FLASH at DESY Hamburg,  a drive laser with the pulse repetition rate of $1$ MHz or $3$ MHz, wavelength of $262$ nm, and pulse energy of $30~\mu$J has been developed~\cite{{WillI2011}}. 
 
DC-SRF photoinjector, which combines a DC pierce gun and a superconducting radiofrequency (SRF) cavity, was first proposed by Peking University in 2001~\cite{ZhaoK2001}. It is a potential candidate of SRF photoinjector to provide electron beam  with medium average current, low emittance, and short bunch length. A prototype of this kind of injector with a 1.5-cell SRF cavity was designed and constructed in 2004~\cite{HaoJ2006}. The preliminary experiment at $4$ K demonstrated the feasibility of DC-SRF structure. An upgraded DC-SRF photoinjector with a 3.5-cell large-grain SRF cavity has been designed and constructed recently and the beam experiment was carried out. The results indicated that the drive laser system used for the 1.5-cell prototype~\cite{LuX2004} is not stable enough to obtain high-quality electron beam. The main problems are low laser power at $1064$ nm and  instability of second harmonic generation (SHG) and fourth harmonic generation (FHG) induced by thermal effects. We have therefore designed and constructed a new drive laser system, which composes of a $1064$-nm laser oscillator, a four-stage amplifier, harmonic generation system, laser beam transport optics, etc, to meet the requirements of the 3.5-cell DC-SRF photoinjector. In this paper, we introduce this new drive laser system and discuss its special features in detail.

\section{Basic parameters and layout of the drive laser system}

The main parameters of the DC-SRF photoinjector are shown in Table~\ref{dcsrfpar}. Considering the stability and the less stringent requirement to vacuum environment, Cs${}_2$Te photocathode is used even though it does not have a very high quantum efficiency and needs to be driven by ultraviolet (UV) laser with the wavelength around $260$ - $270$ nm. 

\begin{center}
\tabcaption{ \label{dcsrfpar}  Main parameters of the DC-SRF photoinjector.}
\footnotesize
\begin{tabular*}{80mm}{l@{\extracolsep{\fill}}cc}
\toprule Parameter & Value & Unit\\
\hline
RF frequency                & $1300$       & MHz \\
Bunch charge                & $>~12$       & pC \\
Average current             & $>~1$        & mA \\
Normalized emittance        & $\sim 1.2$   & $\mu$m \\
Bunch length (FWHM)         & $\sim 3$     & ps \\
Beam energy                 & $3-5$        & MeV \\
\bottomrule
\end{tabular*}
\vspace{0mm}
\end{center}
\vspace{0mm}

The primary parameters of drive laser can be derived from the requirements of the DC-SRF photoinjector. The average current and bunch charge shown in Table 1 indicate that the drive laser does not need to have a repetition rate as high as $1.3$ GHz. We have chosen a repetition rate of $81.25$ MHz, the 16th harmonic of the RF frequency of the 3.5-cell SRF cavity. Since most commercially available laser oscillators work in infrared (IR) wavelength region, the laser oscillator with the wavelength of $1064$ nm, which was used for the 1.5-cell DC-SRF photoinjector prototype, is still adopted for the new drive laser system. Nonlinear crystals will be used to convert the IR pulses to UV pulses with a wavelength of $266$ nm. 

The output power is another important parameter of drive laser system. The photocurrent can be described with the formula~\cite{MontgomeryEJ2007}
\begin{equation}
\label{eq1}
I \textrm{[mA]} = \frac{\lambda\textrm{[nm]}}{124}\times P_{\textrm{laser}}\textrm{[W]}\times Q.E.\textrm{[\%]}.
\end{equation}
where $\lambda$ is the wavelength of drive laser, $Q.E.$ is the quantum efficiency of photocathode. The $Q.E.$ of Cs${}_2$Te photocathode has a maximum of around $10\%$, but it is very sensitive to vacuum environment. Considering a safety margin of $10$, a UV laser with the average power of at least $0.5$ W is needed to obtain an average electron beam current of more than $1$ mA. The expected electron bunch length from DC-SRF photoinjector is around $3$ ps (FWHM). According to the simulation of bunch compression by the RF field in the 3.5-cell SRF cavity~\cite{ZhuF2011}, the drive laser is required to have a pulse length of less than $6$ ps (FWHM).

An optical layout of the drive laser system is shown in Fig.~\ref{lasersystemlayout} and a photo of the laser system can be seen in Fig.~\ref{lasersystemphoto}. The system consists of a $1064$-nm laser oscillator, a four-stage amplifier, the harmonic generation system, the optical system to transfer the UV pulses to the photocathode, and the synchronization system. The laser oscillator produces seed pulses with a repetition rate of $81.25$ MHz. These pulses are synchronized to the RF field in the linac cavities with sub-picosecond timing jitter. The $1064$-nm seed laser pulse from the oscillator is amplified by about $8$ times in power and then converted to $532$-nm green laser and $266$-nm UV laser by the harmonic generators. The UV pulses are finally transported onto the cathode of the DC-SRF photoinjector through proper imaging optics. The details of the design considerations and the performance of laser system will be described in the following sections. 

\end{multicols}
\begin{center}
\includegraphics[width=15cm]{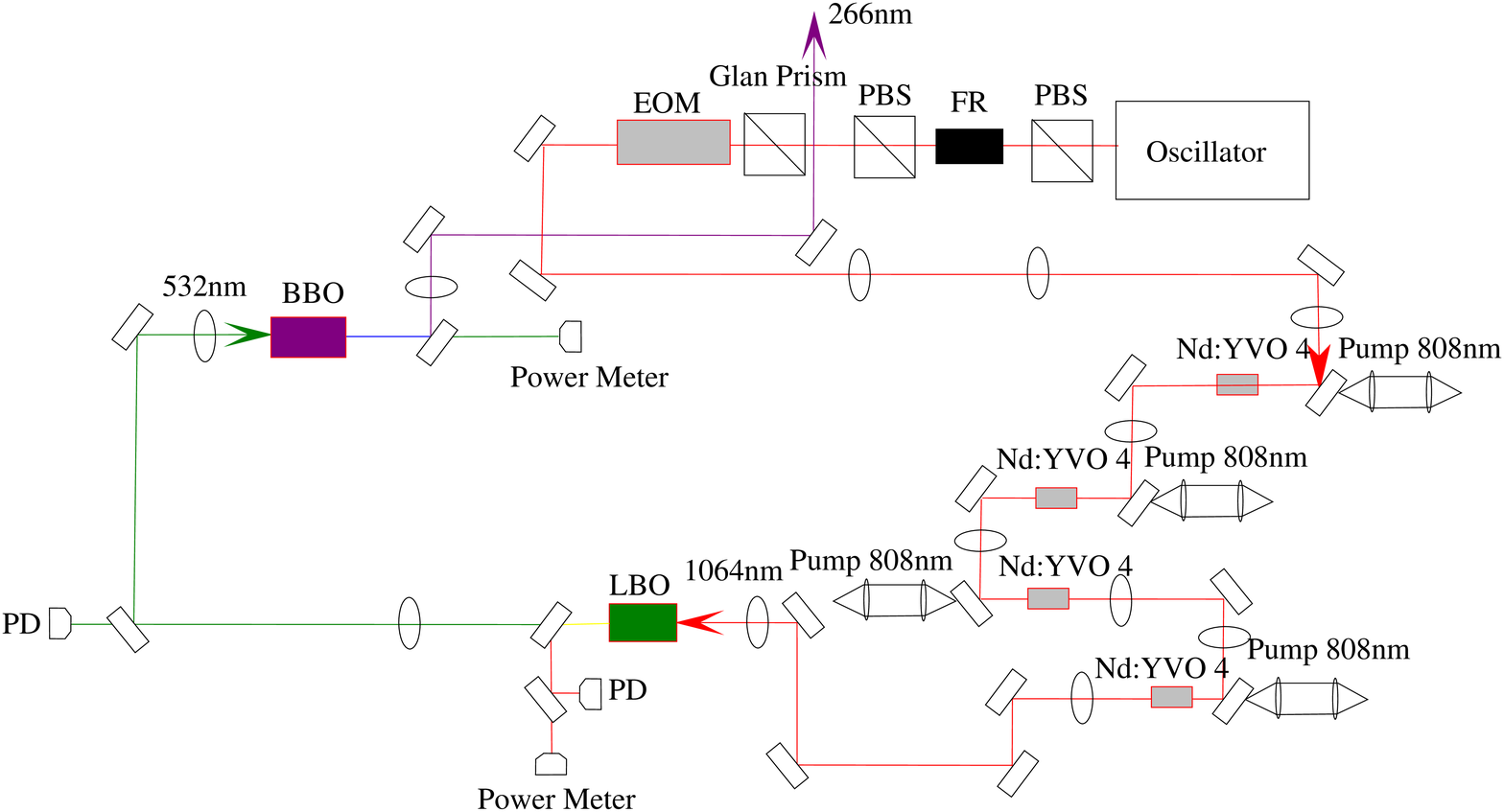}
\figcaption{\label{lasersystemlayout} Schematic layout of the drive laser system for the DC-SRF photoinjector.}
\end{center}

\begin{multicols}{2}

\begin{center}
\includegraphics[width=8cm]{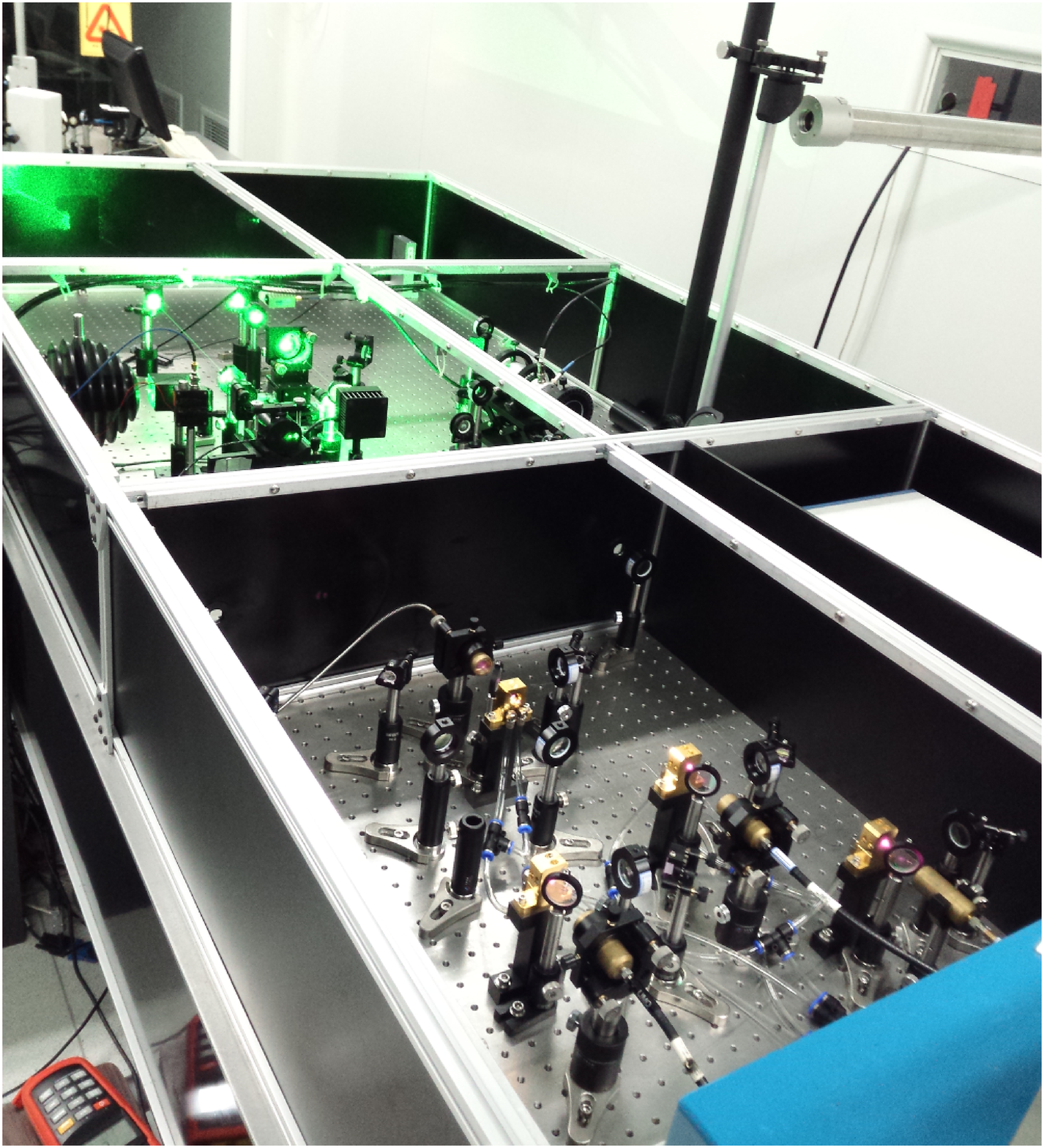}
\figcaption{\label{lasersystemphoto}   A photo of the drive laser system for the DC-SRF photoinjector. The laser system is shielded in a box made of $3$-mm thick aluminum plate.}
\end{center}

\section{Laser oscillator and amplifier}

The $1064$-nm laser is configured as a master oscillator power amplifier (MOPA). The master oscillator, Time-Bandwidth Products GE-100-XHP series, is a passively mode-locked, diode-pumped solid-state laser. The laser pulse from the oscillator has a repetition rate of $81.25$ MHz and a nearly Gaussian temporal profile with a FWHM of $10$ ps. The pulse forming process is started and stabilized by a semiconductor saturable absorber mirror (SESAM), which also forms one end of the laser cavity. It is mounted on a translation stage driven by a picomotor, through which the cavity length can be adjusted. The cavity can also be fine-tuned using a piezo. A  photodiode with nanosecond-level response is applied to monitor the laser pulses. Its signal is phase-locked to the accelerator RF using Time-Bandwidth CLX-1100 timing stabilizer. 

To obtain more stable UV laser pulses, we prefer to use shorter nonlinear crystals for wavelength conversion. To compensate the reduction of wavelength conversion efficiency, much higher $1064$-nm laser power is demanded. A $1064$-nm laser amplifier should therefore be applied. Nd-doped laser materials are widely used for amplifies in many laboratories, such as FLASH, Daresbury, PITZ, FZD Rossendorf, Fermilab, SLAC, etc., due to its high efficiency, good reliability, relatively simply setup, and low maintenance costs. Moreover, low doping concentration Nd:YVO4 crystal has the advantage in reducing thermal loading density and achieving more uniform absorption under high intensity pumping. As a result, a-cut $0.3\%$ Nd-ion concentration bulk Nd:YVO4 crystals ($3$ mm $\times$ $3$ mm $\times$ $10$ mm) are used for the amplifier. Both end faces of the Nd:YVO4 crystals are dichroic coated for high transmittance at $1064$ nm and $808$ nm. The crystals are wrapped with indium foil and placed in copper heat sinks with water cooling. 

For high power output and good beam quality,  the amplifier is configured as four single-pass, single-end-pumped stages. Because of the high repetition rate of the laser pulses, CW laser diode modules need to be used as pump source to keep the laser power instability below $1\%$.  Four sets of LIMO $808$-nm off-board laser diode modules, with the output power of $37$ W, are therefore chosen as the pump sources.  The pump lasers, transported through $400~\mu$m fibers, are coupled onto the end face of the Nd:YVO4 crystals using imaging optics. The pump source couplers are placed in holders which can realize four-dimensional translation and tilt adjustment. It is convenient for optimization and maintenance of the amplifier.

To direct the $1064$-nm laser through the Nd:YVO4 crystals,  four dichroic mirrors, anti-reflection (AR) coated at $808$ nm and high-reflection (HR) coated at $1064$ nm, are placed between the pump source couplers and crystals with the incident angle of $45^\circ$ (see Fig. 1). Telescope optics are both used to match the $1064$-nm laser beam into the Nd:YVO4 crystals and to expand the laser beam when necessary to avoid the damage of optical elements due to too high laser power density. To prevent the amplified laser beam from reflecting back into the master oscillator, the amplifier is isolated from the oscillator using a Faraday rotator.

\begin{center}
\includegraphics[width=8cm]{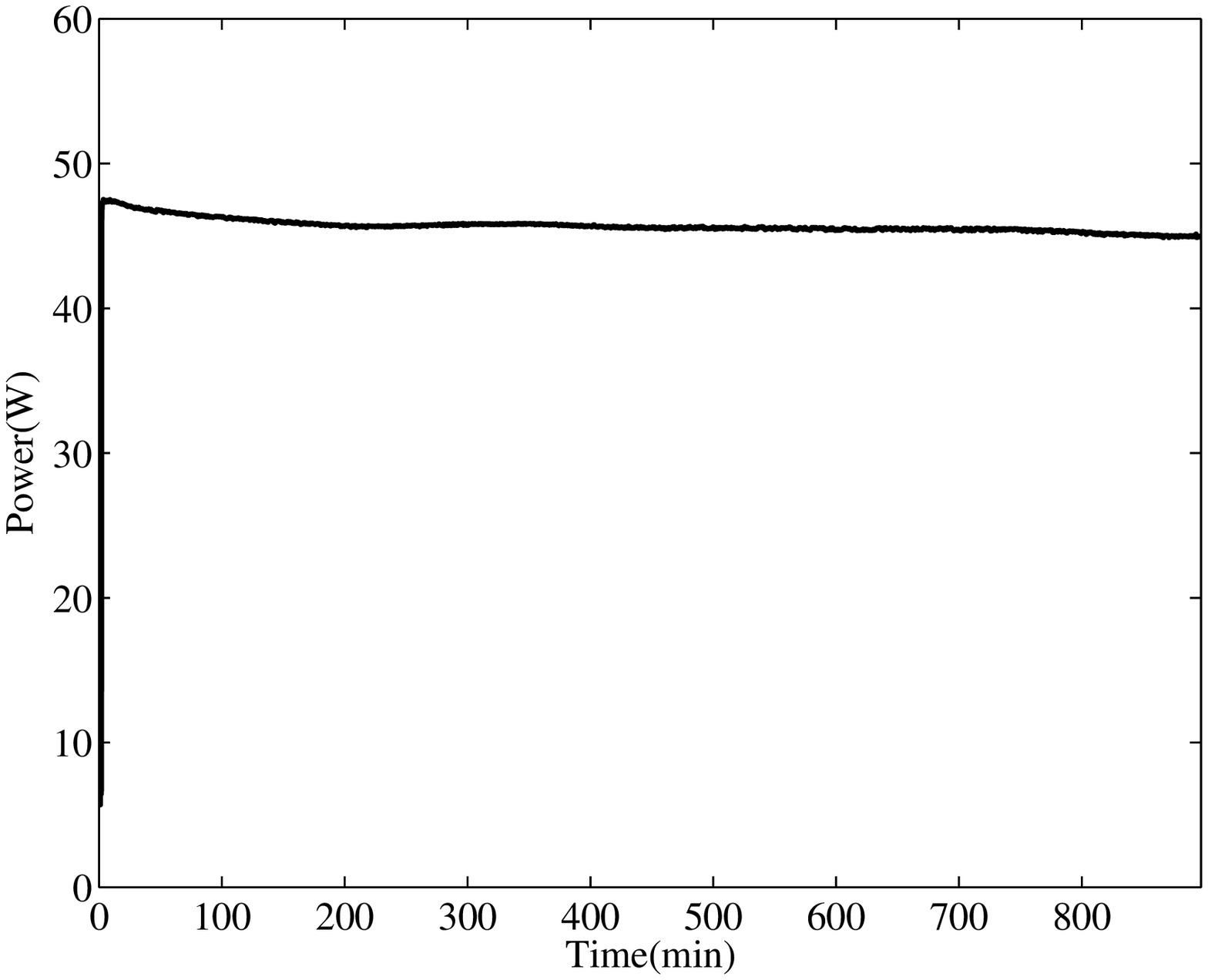}
\figcaption{\label{pow1064}   $1064$-nm laser power after the four-stage amplifier during a long-term monitoring. }
\end{center}

The $1064$-nm laser power is $5$ W before the first stage of the amplifier. In CW operation, at a pump current of $45$ A, $45$ W stable output has been obtained from the MOPA system (see Fig.~\ref{pow1064}). The net power gain of the amplifier is $40$ W. Compared to the total pump power of the four laser diode modules ($148$ W), the total optical-optical efficiency is about $27\%$. Fig.~\ref{pow1064vscur} shows the average laser power after the amplifier versus the current of the pump diode power supplies. It can be seen from the figure that the average laser power grows rapidly as the current increases above $25$ A. The increase of the laser power slows down around $42$ - $45$ A. The operation pump current is therefore set at $45$ A, which is still within the safe region.

\begin{center}
\includegraphics[width=8cm]{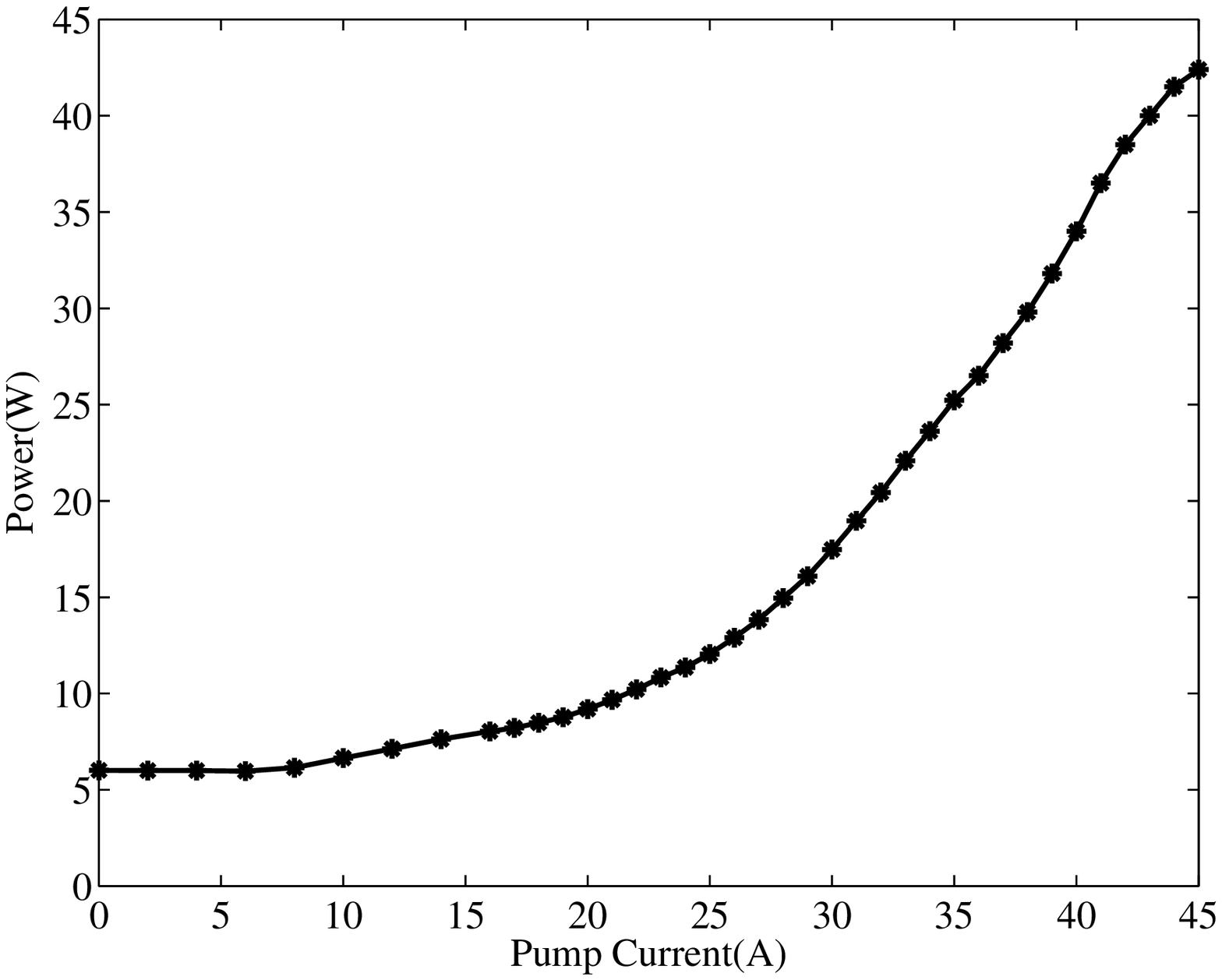}
\figcaption{\label{pow1064vscur}   Amplified $1064$-nm laser power as a function of the amplifier pump diodes's current. }
\end{center}

\section{Harmonic generation system}

To obtain the UV laser pulses required by the Cs${}_2$Te cathode in the DC-SRF photoinjector, a two-stage cascaded harmonic generation system has been designed. As the first stage, a LBO crystal is used for SHG. The $532$-nm laser from LBO then traverses a BBO crystal for FHG. The LBO and BBO crystals are chosen for SHG and FHG because they are non-critically phase-matched and have high damage threshold and high conversion efficiency.

The LBO crystal has a length of $15$ mm and a cross section of $4$ mm $\times$ $4$ mm and is cut at $\theta$ of $90^\circ$ and $\varphi$ of $10.6^\circ$. Its end faces are AR coated for high transmittance ($T > 99.8\%$) at $1064$ nm. The crystal is mounted in a thermostat, whose temperature can be remotely adjusted for optimized phase-matching angle. During the operation, the temperature is controlled at $45\pm0.1^\circ C$.

For FHG using BBO crystal, the large walk-off angle~\cite{BoydG1968} caused by BBO crystal's large birefringence limits the conversion efficiency and makes the transverse profile of UV laser beam far from Gaussian distribution. This problem can be solved by using a walk-off compensation system, where two or more crystals having the same cut and length~\cite{DrozC,FriebeJ,OndaT2013} are arranged with the crystal optic axes in alternating directions (see Fig.~\ref{walkoffcompen}).  In our system, a pair of $2$-mm long BBO crystals, which have a $4$ mm $\times$ $4$ mm cross section and are cut at $\theta$ of $47.7^\circ$ and $\varphi$ of $0^\circ$, are used. The BBO crystals' end faces are AR-coated for high transmittance at $532$ nm ($T > 99.8\%$ ) and $266$ nm ($T > 99.5\%$). 

\begin{center}
\includegraphics[width=8.5cm]{fig_2bbos.eps}
\figcaption{\label{walkoffcompen}   Walk-off compensation arrangement using two BBO crystals for second harmonic generation from green laser to UV laser. The extraordinary UV beam walk-off in the first crystal is compensated by propagation through the second crystal, where the extraordinary UV beam walks on to the two ordinary beams from the two crystals.}
\end{center}

As shown in Fig.~\ref{lasersystemlayout}, demagnifying telescopes located before the LBO and BBO crystals are used to reduce the diameter of the beam to $10s~\mu$m for better wavelength conversion performance. After the LBO crystal, a set of lens and dichroic mirror is employed to collimate the SHG beam and eliminate the residual $1064$ nm laser. After the BBO crystals, another set of lens and dichroic mirror is also employed to collimate the FHG beam and eliminate the residual $532$ nm laser.

\begin{center}
\includegraphics[width=8.5cm]{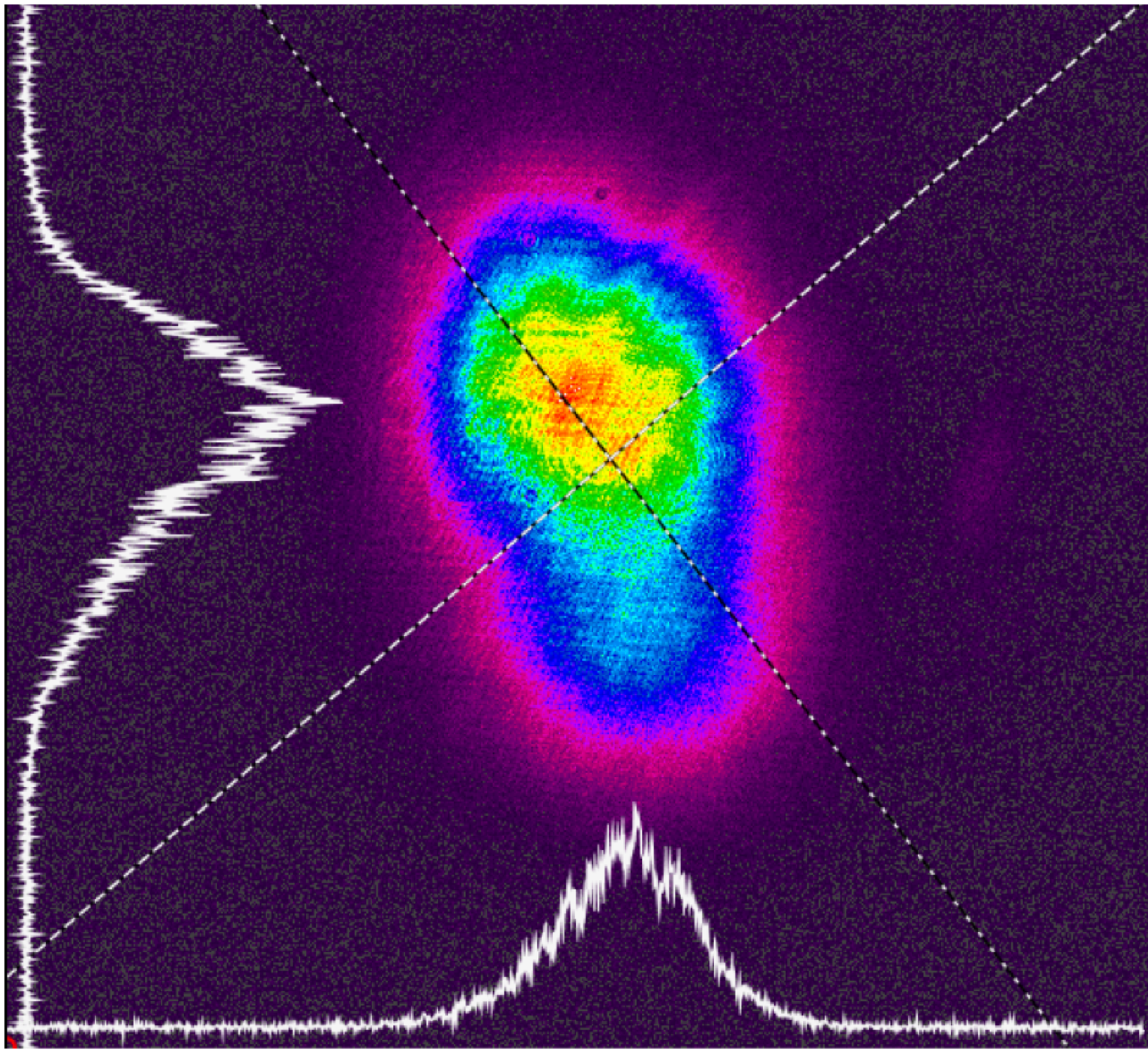}
\figcaption{\label{laser266profl}   Two-dimensional and three-dimensional transverse profiles of the $266$-nm UV laser. }
\end{center}

The measured conversion efficiencies of the SHG and FHG  are $40\%$ and $5.6\%$, respectively. The total conversion efficiency, from $1064$ nm to $266$ nm, is better than $2\%$. The $266$-nm UV laser beam profile has been improved compared to our previous laser system using single-BBO (see Fig.~\ref{laser266profl}). It can be further improved using transverse distribution shaping techniques. There is still enough room for increasing the conversion efficiency by different design of the harmonic generation system. Especially, an eletro-optic (EO) modulator is being used to adjust the repetition rate of the $1064$-nm laser pulses. Using the EO modulator, the laser pulse repetition rate can be reduced as desired. This will benefit the harmonic generation system, since the laser pulse energy can be increased while the average power remains unchanged.

\section{Stability improvement of the drive laser system}

The laser system is installed in a class 1000 cleanroom, where the temperature is controlled at $20 \pm 0.5^\circ C$ and humidity less than $30\%$. Because of the air conditioning system, air flow exists in this cleanroom. To reduce the vibration caused by air flow, the main parts of the drive laser system is covered in a shielding box made of $3$ mm thick aluminum plate (see Fig.~\ref{lasersystemphoto}). 

In the previous drive laser system, the FHG system was installed in our SRF accelerator hall and the instabilities of both laser power and laser beam pointing became enlarged. In our new design, the FHG has been moved into the cleanroom. This makes the $266$-nm laser transporting a longer distance (about $20$ m) to the reflecting mirror installed in the vacuum chamber of electron beam line (see Fig.~\ref{uvtransport}), and it is more difficult to project the laser beam to the photocathode. A motorized mirror mount has been designed and installed in the optical beam line for precise remote control of a $45^\circ$ mirror right before the in-vacuum reflecting mirror. The incident point of UV laser on the surface of the photocathode can therefore be precisely scanned using this motorized mirror mount. The optical beam line is shielded to prevent air turbulence caused by a temperature differential and air flow.

\begin{center}
\includegraphics[width=8cm]{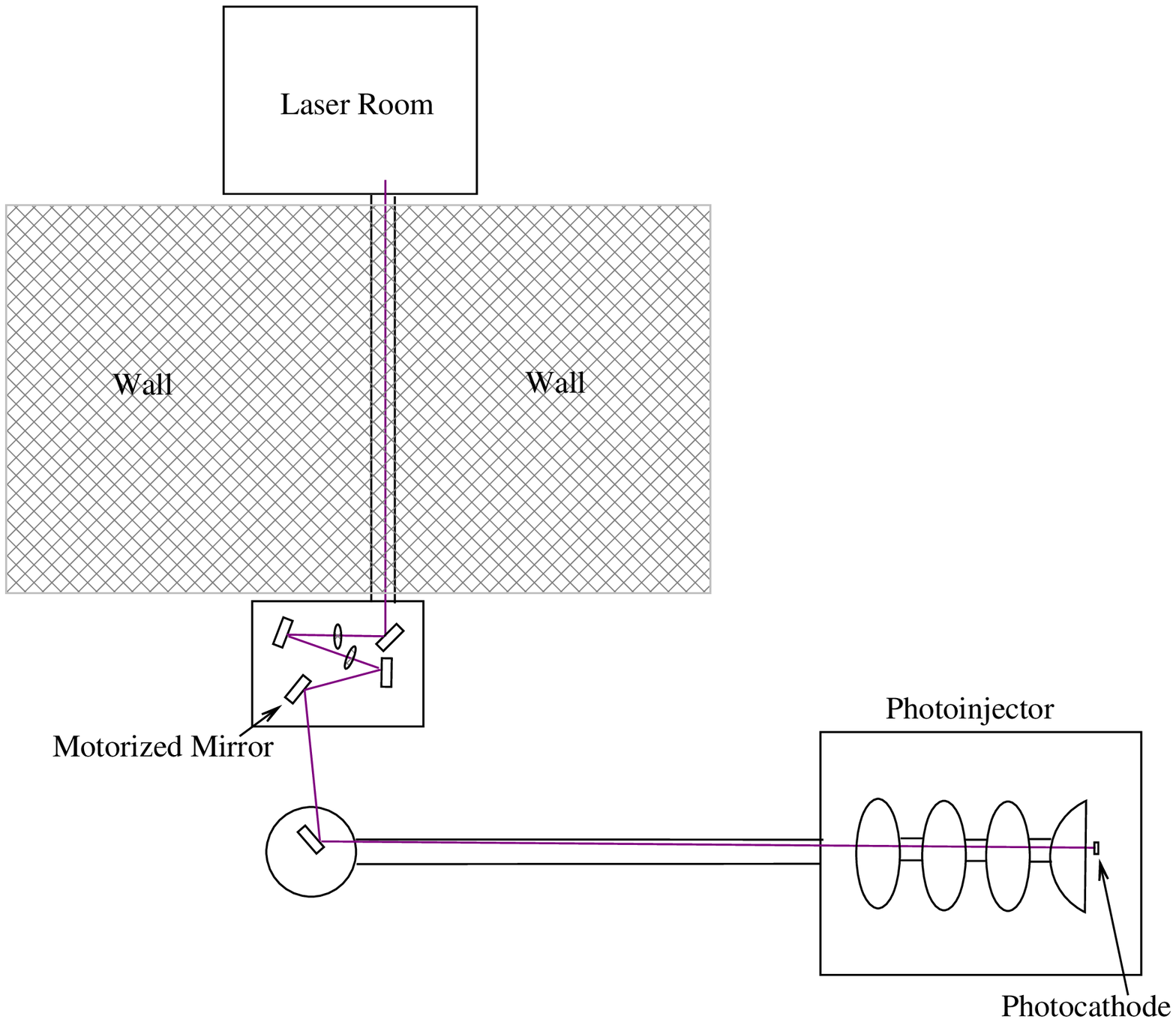}
\figcaption{\label{uvtransport}   Optical beam line for $266$-nm UV laser. }
\end{center}

\section{Performance of the drive laser}

The performance of the drive laser system has been investigated, including the power instability, the beam pointing instability, and the synchronization with accelerator RF signal. The long-term instability of the $1064$-nm and $266$-nm lasers were monitored using a power meter integrated into our control system. Fig.~\ref{pow1064} and Fig.~\ref{pow266} present the results of our latest monitoring of the $1064$-nm laser and the $266$-nm laser, respectively. During the measurements, the drive laser system was operated with the $1064$-nm laser power of $45$ W and the $266$-nm laser power close to $1$ W. The long-term instability of UV laser power is less than $5\%$ after the warming up of the laser system. It is sufficient for the DC-SRF photoinjector beam experiments. 

\begin{center}
\includegraphics[width=8cm]{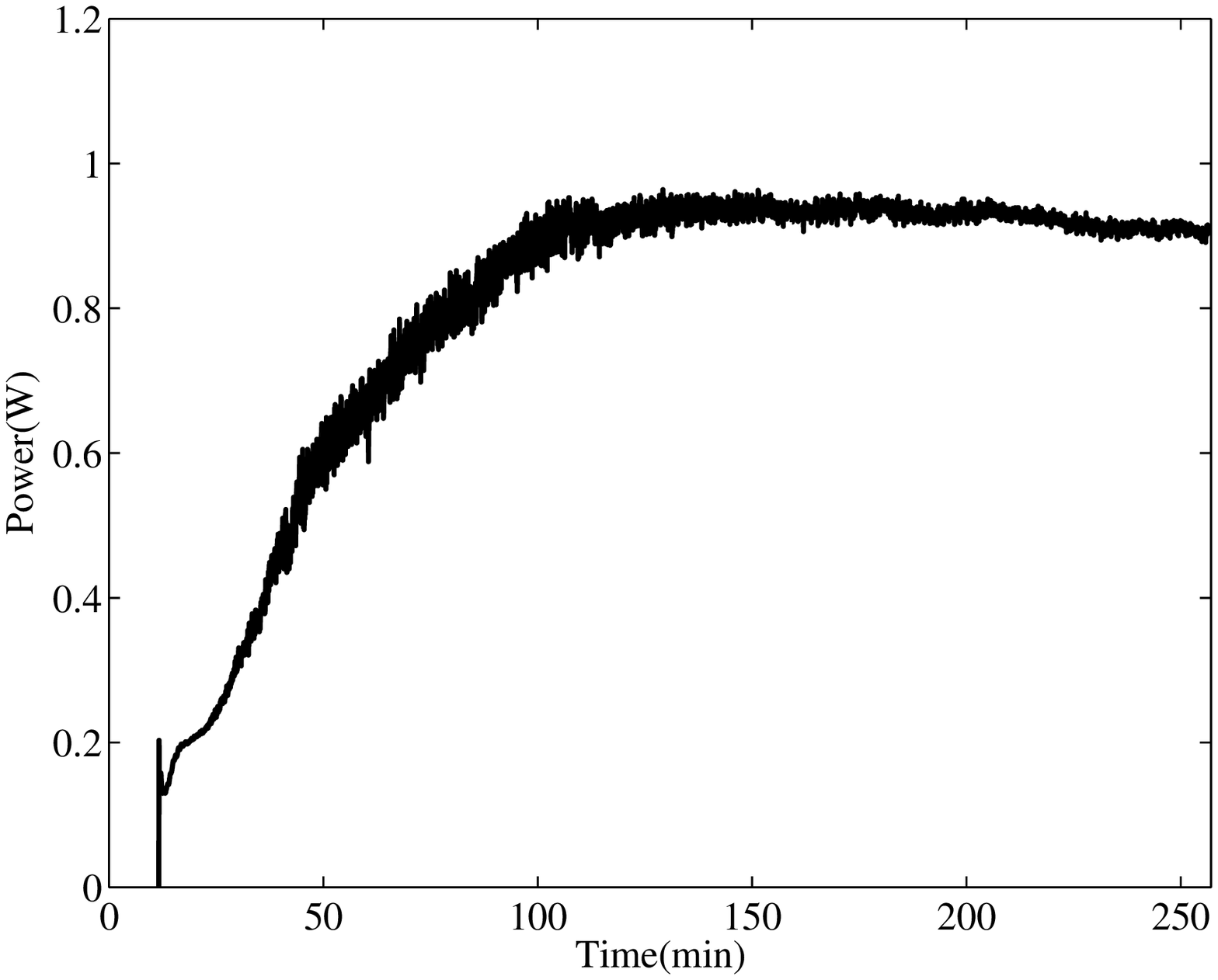}
\figcaption{\label{pow266}   $266$-nm laser power from warming up to stable operation. }
\end{center}

The pointing stability of the $266$-nm laser beam was estimated by analyzing the laser beam on a virtual photocathode using a beam profiler. The pointing instability was estimated to be $10~\mu$rad rms in both horizontal and vertical direction. It is remarkably improved compared to our previous drive laser system. 

A phase detector, integrated with a fast photodiode, was used to monitor the phase difference between the $1064$-nm laser pulses and accelerator RF field. The timing jitter, derived from the phase difference, was estimated to be less than $1$~ps rms. The phase difference between the $266$-nm laser pulses and accelerator RF field was measured by a UV phase detector. It is also less than $1$~ps rms. This means that the UV laser beam is well isolated from environmental vibrations. 

The new drive laser system has been applied in the beam test and stable operation of the DC-SRF photoinjector. With $0.5$~W $266$-nm laser power, a stable electron beam has been obtained with an energy of $3.4$~MeV. The average beam current reached $1$ mA and was kept at $0.55$~mA for long-term operation. The electron beam current is not limited by the drive laser system at present. 

\section{Conclusion}

A new drive laser system has been designed and constructed for the upgraded 3.5-cell DC-SRF photoinjector at Peking University successfully. Double BBO crystals in the FHG are adopted to reduce the walk-off effect. Careful design makes the system more compact and isolated from interferences of surrounding environment. The $266$ nm laser with the average power close to $1$ W can be delivered to illuminate the Cs${}_2$Te photocathode and the instability is less than $5\%$ for long-term operation. The pointing instability was estimated to be $10~\mu$ rad rms in both horizontal and vertical direction. The drive laser system has been applied in the DC-SRF photoinjector beam experiments and the performance meets the requirements. Recently, an EO modulator has been installed in our laser system. Using the EO modulator, the laser pulse repetition rate can be reduced as desired. The laser system, especially the harmonic generators, may be further developed to make the best use of this new capability.

\acknowledgments{}

\end{multicols}

\vspace{15mm}

\vspace{-1mm}
\centerline{\rule{80mm}{0.1pt}}
\vspace{2mm}

\begin{multicols}{2}

\end{multicols}

\clearpage

\end{document}